\documentclass[twocolumn,preprintnumbers,amssymb,amsmath]{revtex4}

\usepackage{graphicx}% Include figure files
\usepackage{dcolumn}% Align table columns on decimal point
\usepackage{bm}% bold math

\catcode`\@=11
\def\lsim{\mathrel{\mathpalette\@versim<}}
\def\gsim{\mathrel{\mathpalette\@versim>}}
\def\@versim#1#2{\vcenter{\offinterlineskip
\ialign{$\m@th#1\hfil##\hfil$\crcr#2\crcr\sim\crcr } }}
\catcode`\@=12

\newcommand{\nn}{\nonumber}
\newcommand{\bi}{\bibitem}
\newcommand{\be}{\begin{equation}}
\newcommand{\ee}{\end{equation}}
\newcommand{\bea}{\begin{eqnarray}}
\newcommand{\eea}{\end{eqnarray}}

\begin{document}
\input epsf.tex

%\tightenlines
%\draft

\preprint{KUNS-1949, NIIG-DP-04-3, KANAZAWA-04-20}
\title{Induced top Yukawa coupling and 
suppressed Higgs mass parameters}
\author{Tatsuo Kobayashi}
\email{kobayash@gauge.scphys.kyoto-u.ac.jp}
\affiliation{
Department of Physics, Kyoto University, 
Kyoto 606-8502, Japan
}%
\author{Hiroaki Nakano}
\email{nakano@muse.sc.niigata-u.ac.jp}
\affiliation{
Department of Physics, Niigata University, Niigata 950-2181, Japan
}%
\author{Haruhiko Terao}
\email{terao@hep.s.kanazawa-u.ac.jp}
\affiliation{
Institute for Theoretical Physics, Kanazawa
University, Kanazawa 920-1192, Japan
}%
\date{\today}
\begin{abstract}
In the scenarios with heavy top squarks,
mass parameters of the Higgs field
must be fine-tuned due to a large logarithmic 
correction to the soft scalar mass.
We consider a new possibility that the top 
Yukawa coupling is small above TeV scale. 
The large top mass is 
induced from strong Yukawa interaction of 
the Higgs with another gauge sector, in which 
supersymmetry breaking parameters are given
to be small.
Then it is found that the logarithmic correction
to the Higgs soft scalar mass is suppressed
in spite of the strong coupling and the fine-tuning
is ameliorated.
We propose an explicit model coupled to a 
superconformal gauge theory which realizes
the above situation.

\end{abstract}

%\keywords{MSSM, Little hierarchy, Superconformal, Top mass}

%\narrowtext

\maketitle

%%%%%%%%%%%%%%%%%%%%%%%%%%%%%%
\section{Introduction}
Supersymmetric little hierarchy \cite{BG,finetune} means a 
large discrepancy
between the scale of weak boson masses and the scale of 
supersymmetry breaking masses, specially for the Higgs fields.
In the minimal supersymmetric standard model (MSSM),
the minimization condition for the potential of 
neutral Higgs components is given by
\be
\frac{M_z^2}{2} = -\mu^2 - m_{H_u}^2,
\label{minimization}
\ee
for a moderate value of $\tan \beta$.
Here $m_{H_u}$ denotes the soft supersymmetry breaking
mass for the up-type Higgs field $H_u$, while 
$\mu$ denotes the supersymmetric mass parameter.
If $|m_{H_u}|$ is much larger than $M_z$, the two
mass parameters $m_{H_u}$ and $\mu$ must be fine-tuned
so as to nearly cancel out each other.
However there is no theoretical ground that low energy
values of these parameters are related mutually.
Therefore, it seems to be natural for the Higgs mass 
parameters, $|m_{H_u}|$ and $|\mu|$, to appear less 
than a few hundred GeV.

Contrary to this, the soft supersymmetry breaking mass
$|m_{H_u}|$ appears to be fairly large  at low energy
in the most supersymmetric models.
This is due to the large radiative correction,
which is  given explicitly at one-loop level by
\be
\delta m_{H_u}^2 \sim - \frac{12 }{16\pi^2}Y_t^2
m_{\tilde t}^2 \ln \frac{\Lambda}{m_{\tilde t}}.
\label{log-h-mass}
\ee
Here $Y_t$ and $m_{\tilde t}$ denote the top Yukawa 
coupling and the soft supersymmetry breaking mass 
of stop (top sparticle).
We represent an uppermost scale of the MSSM by $\Lambda$,
which may be taken to be the GUT scale; 
$\Lambda \simeq 10^{16}$GeV.
Then the correction $|\delta m_{H_u}|$ is found to be
comparatively larger than $m_{\tilde t}$ because of the
sizable top Yukawa coupling $Y_t$.
This negative correction to the Higgs scalar mass squared
gives rise to radiative Electro-Weak symmetry breaking (EWSB)
\cite{EWSB},
which is thought to be one of the beautiful features of the
MSSM.
However the problem is that the radiative correction is 
too large, since $m_{\tilde t}$ is supposed to be more than
several hundred GeV for the following reasons.

First one is based on the experimental lower bound of 
the lightest Higgs boson mass $m_{h^0}$, which is 
114 GeV \cite{LEPII}.
On the other hand, the MSSM predicts the lightest CP-even Higgs 
boson mass $m_{h^0}$ to be less than $M_Z$ at the tree level.
This discrepancy can be covered by sizable radiative 
correction for $m_{h^0}$, when the top-stop mass splitting is 
relatively large \cite{massbound}.
Explicitly, the one-loop correction may be written roughly as 
\begin{equation}
\Delta m_{h^0}^2 \sim \frac{3}{4\pi^2}Y_t^2 m_t^2
\ln \left(\frac{m^2_{\tilde t}}{m_t^2}\right),
\label{h-mass}
\end{equation}
where $m_t$ denotes the top mass. 
This formula requires $m_{\tilde t} \geq 500$GeV
so as for the lightest Higgs mass in the MSSM
to satisfy the experimental bound.
Then the soft supersymmetry breaking mass $|m_{H_u}|$
is supposed to be more than about $800$GeV in the MSSM, 
and rather tight fine-tuning less  than a few percent
is required in order to satisfy Eq.~(\ref{minimization}).

Second one is theoretical. One of strong supports for
the low energy supersymmetry is gauge coupling 
unification around the scale 
$\Lambda \sim 2 \times 10^{16}$GeV.
When the gaugino masses $M_a (a=1,2,3)$ are also unified 
around $\Lambda$, the gaugino masses enjoy the famous 
GUT relation with the corresponding gauge coupling $g_a$ as 
$M_a/g^2_a = {\rm const}$.
Therefore, the gauge coupling unification leads to a rather
heavy gluino mass at low energy.
This is the same for the minimal supergravity (mSUGRA) model
\cite{finetune}. 
Such a gluino mass enhances the soft supersymmetry 
breaking masses of squarks at low energy through 
radiative corrections. Explicitly the
correction to $m^2_{\tilde{t}}$ becomes as large as 
$(500 {\rm GeV})^2$,
even if $M_1 \sim 100$GeV at low energy.
Then the soft mass of Higgs receives large radiative
corrections through these heavy squarks.
Thus we may say that the primary origin of the fine-tuning
problem is the enhanced gluino mass.

The same fine-tuning problem arises in the models with
the so-called gauge mediated supersymmetry breaking (GMSB),
which is one of the most postulated scenarios explaining
the flavor universal soft masses.
There the soft masses and also the gaugino masses are
generated at some lower scale than the GUT scale.
However the size of the generated squarks masses as well as
the gaugino masses are found to be large as those 
given in the mSUGRA scenario.
Thus the fine-tuning problem is found to be rather severe
also in the GMSB \cite{GMSB}.

There have been various proposals to remedy this fine-tuning 
problem by considering extensions of the MSSM.
Many of them are to enhance the effective quartic coupling 
of the Higgs field in the low energy theory. 
The ways are various; 
low energy supersymmetry breaking \cite{Casas}, 
additional F-term contributions \cite{fathiggs}, 
additional D-term contributions \cite{Delgado}, 
an additional sizable Yukawa coupling of the Higgs 
field with extra fields \cite{Babu}.
Indeed enhancement of the quartic coupling may improve 
degree of the fine-tuning somewhat, 
however this effect is not so significant. 
Rather it is essential that the tree level value of the
lightest Higgs boson mass is enhanced and the large
stop mass is not required any more to satisfy the
experimental bound.
Then the radiative correction (\ref{log-h-mass})
to the Higgs soft mass
may be sufficiently reduced in the case of light stop masses.
Thus it should be noted that light gluino mass as well as
the light stop mass must be assumed in these scenarios,
which does not conform to the sparticle spectra obtained
in the frameworks of mSUGRA nor GMSB.

Therefore it would be more interesting to study
the models in which  
only the Higgs mass parameters are suppressed with keeping
the other soft masses heavy, as given in the GMSB scenario.
In order to realize such a situation, it is necessary to eliminate
or suppress 
the logarithmic correction to the Higgs soft mass, 
$\delta m_{H_u}^2$, which is given by 
Eq.~(\ref{log-h-mass}) approximately.
Actually there have been proposed only a few scenarios along this 
line of thought. 
One way is given by the so-called supersoft
supersymmetry breaking mechanism \cite{supersoft}.
There the soft masses are generated only
through finite loop diagrams so that the flavor universality
of them is automatically guaranteed.
On top of that, the soft scalar mass of Higgs
is free from logarithmic enhancement
and is also suppressed by one-loop factor.
Another one is the supersymmetric little Higgs model 
\cite{superlittle}, in which the Higgs fields are assumed to 
be pseudo Nambu-Goldstone bosons,
and their mass parameters are generated only radiatively.
Then the enlarged symmetry ensures that the logarithmic 
corrections to the mass parameter
appear at more than two-loop level and, therefore, are
suppressed.

Two of the authors also presented models in which the Higgs
field has an additional Yukawa coupling with 
a superconformal gauge sector \cite{KT}.
It has been known \cite{Karch,sumrule} that soft masses
introduced to the chiral matter fields of a superconformal gauge
theory enjoy sum rules at infrared (IR) irrespectively of their
values at high energy scale.
In the previous paper, it was shown that the Higgs soft mass 
can be made finite and ultraviolet insensitive
by using the sum rules.
Moreover, the obtained Higgs mass is one-loop suppressed.
Besides the Higgs field acquires a large anomalous 
dimension and, therefore, the $\mu$ parameter is suppressed
at the same time.
Thus both of the mass parameters of the Higgs field 
turn out to be
small at low energy by the same dynamics and become free from
fine-tuning. 
However the explicit model relies upon 
assumptions to unknown dynamics
in order to realize the finite soft masses and may be 
somewhat artificial.
Besides that, it is an undesirable feature that top Yukawa coupling
is suppressed at low energy due to the large anomalous dimension 
of the Higgs field
\footnote{Recently Luty and Okui \cite{Luty} also considered 
suppression of the Higgs mass parameters by use of conformal 
dynamics in non-supersymmetric case in order to solve 
the hierarchy problem. 
There also the top Yukawa coupling appears to be suppressed 
at low energy.
}.

In this paper we seek for another possibility that the Higgs
field in the MSSM is sequestered from gluino and top squarks,
whose large masses cause the fine-tuning problem.
We suppose that the top Yukawa coupling is small above
TeV scale.
Then the radiative correction given in Eq.~(\ref{log-h-mass})
is suppressed and becomes harmless.
This radical assumption also requires a new source for the 
top mass in place of the top Yukawa coupling.
So we introduce a strongly coupled gauge sector and a new
Yukawa interaction between the Higgs field and the matter
fields in the strongly coupled sector.
The large top mass is found to be induced at low energy
through mixing between the fundamental top quarks and the 
strongly coupled matter fields. 
One might wonder that the new Yukawa coupling induces a large
correction to the Higgs mass parameter just as the top
Yukawa coupling does, and is not helping to solve the problem.
However if the supersymmetry breaking parameters in the
strong gauge sector are sufficiently small for some 
reason, then the Higgs mass is not enhanced through the
new Yukawa coupling.
Here it is noted that if the strong gauge theory is
superconformal, then the gaugino mass and the A-parameters
necessarily vanish.
Later we will give an explicit model in which mass
of the Higgs is preserved to be small due to 
the superconformal dynamics.
We stress that the superconformal dynamics itself
does not suppress the soft scalar mass of Higgs, and
the scenario considered here is distinct from the previous 
one considered in Ref.~\cite{KT}.

The article is organized as follows.
In section 2 we explain the basic mechanism for sequestering
Higgs from the large soft masses of gluino and stops by
assuming small top Yukawa coupling above TeV scale.
We present an explicit model using superconformal dynamics
in section 3. There the mixing between top quarks and the
extra matter fields, which induces large top mass, is also shown.
In section 4 we study some phenomenological aspects of
the model given in section 3.
Finally section 5 is devoted to conclusion and discussion.

%%%%%%%%%%%%%%%%%%%%%%%%%%%%
\section{Sequestering Higgs from large supersymmetry breaking}
First let us discuss renormalization properties of the soft masses 
of Higgs $H_u$ and stop fields $Q_3, \bar{u}_3$ in the MSSM at
one-loop level.
We neglect the bottom Yukawa coupling $Y_b$ and 
the gauge interactions other than $SU(3)_C$, 
as these effects are not significant for the 
fine-tuning problem.
Then the renormalization group (RG) equations for the soft
masses are given by
\bea
16\pi^2 \frac{d m^2_{Q_3}}{d \ln \mu} &=&
 X_t  -\frac{32}{3} g^2_3 |M_3|^2, \nn \\
16\pi^2 \frac{d m^2_{\bar{u}_3}}{d \ln \mu} &=&
2 X_t  -\frac{32}{3} g^2_3 |M_3|^2, \nn \\
16\pi^2 \frac{d m^2_{H_u}}{d \ln \mu} &=&
3 X_t, 
\label{topRG}
\eea
where $X_t$ denotes the following combination of the 
soft scalar masses and the A-parameter $A_t$;
\be
X_t = 2 |Y_t|^2 \left[
\left( m^2_{Q_3} + m^2_{\bar{u}_3} + m^2_{H_u}
\right) + |A_t|^2
\right].
\ee
The parameter $A_t$ follows the RG equation given by
\be
16\pi^2 \frac{d A_t}{d \ln \mu} =
2 |Y_t|^2 A_t  + \frac{32}{3} g^2_3 M_3. 
\ee
The large gluino mass $M_3$ induces large corrections
not only to squark masses $m^2_{Q_3}$ and 
$m^2_{\bar{u}_3}$ but also to $A_t$.
However if the top Yukawa coupling $Y_t$ is given to be small
enough, then the soft mass parameter of 
up-type Higgs $m^2_{H_u}$ is protected from the corrections.

We note also that these RG equations show an interesting
property as follows.
If we drop off $M_3$ and $A_t$ from these equations
and treat $Y_t$ as a constant approximately, 
then these coupled equations are easily solved.
There are two constant modes for 
$(m^2_{Q_3}, m^2_{\bar{u}_3}, m^2_{H_u})$, 
which are proportional to $(1,-1,0)$ and $(1,1,-2)$.
The linearly independent mode proportional to $(1,2,3)$ 
is suppressed towards IR with a power of scale
\footnote{We can always tune the high energy values 
of the scalar masses so that $|m_{H_u}|$ vanishes 
at low energy scale. Such initial values are given
as two parameter solutions.
Accidentally the universal initial values given at GUT
scale happen to lead to vanishing $|m_{H_u}|$ at
the EW scale, which is known as the focus point
\cite{focus}.
However natural explanation for the initial
values is required in turn.
}.
Therefore the soft masses satisfy the sum rule; 
$m^2_{Q_3} + m^2_{\bar{u}_3} + m^2_{H_u} \rightarrow 0$
at low energy.
Here it is noted that the soft masses are not enhanced
from the high energy values.
As long as initial values of the scalar masses
are given to be small, 
$|m_{H_u}|$ is always small at low energy.
Thus the supersymmetric little hierarchy may be
ameliorated by suppressing the gluino mass and the 
A-parameter for the top Yukawa coupling
compared with other gaugino masses.
In this paper we do not pursuit for this possibility
keeping scenarios of mSUGRA and GMSB in our mind.

Now suppose that the top Yukawa coupling $Y_t$
is small {\it e.g.} as $Y_b$. 
Then we need to explain the large top mass by another mechanism.
As was mentioned in section 1, we introduce an extra
strongly interacting gauge theory with the gauge group
$G_S$.
We also assume additional chiral superfields 
$(\Phi, \bar{\phi})$, which are charged under
$G_S$. The quark fields $(Q_3, \bar{u}_3)$ are
charged under $SU(3)'_C$, while $(\Phi, \bar{\phi})$
are singlet.
We consider their Yukawa interaction with $H_u$ as
\be
W \sim Y_t Q_3 \bar{u}_3 H_u + \lambda \Phi \bar{\phi} H_u.
\label{newyukawa}
\ee
The new Yukawa coupling $\lambda$ becomes large
at low energy due to strong gauge interaction.
It is noted that large $\lambda$ reduces  $Y_t$.
In practice we consider scenarios in which the symmetry
$G_S \times SU(3)'_C$ is spontaneously broken to
the color gauge group $SU(3)_C$ at low energy.
After this symmetry breaking, 
$\Phi$ and $\bar{\phi}$
carry the same charges as $Q_3$ and $\bar{u}_3$ have
respectively.
Therefore these extra matter fields can mix with the original
top quarks.
Eventually large top Yukawa
coupling is induced from the second term of the 
superpotential given in (\ref{newyukawa})
in the low energy effective theory.
We will explain how the mixing can be generated
by presenting  an explicit model in the next section.

We also assume that the extra matter fields have 
their vector-like
partners $(\bar{\Phi}, \phi)$ and supersymmetric
mass terms
like $\mu$-term of the Higgs fields;
\be
W \sim \mu H_u H_d + M_{\Phi} \Phi \bar{\Phi} 
+ M_{\phi} \phi \bar{\phi}.
\ee
We will explicitly see later that $\mu$ is 
suppressed, while $M_{\Phi}$ and $M_{\phi}$ are 
enhanced by the strong extra
gauge interaction \cite{KT}. 
Therefore we may suppose that the decoupling
scale given $M_{\Phi}$ or $M_{\phi}$ to be about
several TeV. 
We also assume that the symmetry breaking
takes place around this scale.

How about the radiative correction to the soft mass 
of $H_u$?
What we should care is corrections induced by the
large extra Yukawa coupling $\lambda$.
Here it should be noted that the RG equations for 
soft scalar masses 
$m^2_{\Phi}$, $m^2_{\bar{\phi}}$ and $m^2_{H_u}$
have the same structure as in Eqs.~(\ref{topRG}).
Therefore, if gaugino mass of the extra gauge
theory $M_S$ and also 
the A-parameter for the extra trilinear
interaction $A_{\lambda}$ are both suppressed well,
then the soft scalar mass of Higgs $m^2_{H_u}$
is not enhanced at all irrespectively of the 
strong couplings.
In such a case, the mass parameters of Higgs are
fairly smaller than the gluino mass at
low energy,
as long as $m^2_{\Phi}$ and $m^2_{\bar{\phi}}$
given at high energy scale are also small.
Below the scale of symmetry breaking,
$m^2_{H_u}$ receives the negative
correction through the effective top Yukawa coupling 
just as in the MSSM.
However size of this correction is reduced by 
about one order,
since $\Lambda$ in Eq.~(\ref{log-h-mass}) is now
given to be only several TeV.
Thus radiative EWSB still occurs but the Higgs mass
parameters appearing in Eq.~(\ref{minimization})
are suppressed well.

Now we discuss the ways to realize the above situation.
Indeed there are at least two possibilities in which
the above situation is realized naturally.
The first one is an application of superconformal gauge theories.
We note that both of the gaugino
mass and the A-parameters are suppressed
in proportion to certain powers of the renormalization
scale in any superconformal gauge theories
\cite{Karch,sumrule}.
The superconformal gauge theories are not special
at all. Only if the number of charged matter fields 
is given in the so-called conformal window,
the gauge coupling is found to have an 
IR attractive fixed point \cite{seiberg},
where the theory becomes conformal invariant.
We will give somewhat detailed discussion of this in
section 3.
Thus in the case that the extra gauge sector 
becomes  superconformal with approaching 
an IR fixed point, the Higgs field can be
sequestered from supersymmetry breaking effects.
In the next section we will present an explicit
model suppressing the Higgs mass parameters by
the superconformal dynamics.

Alternatively we may consider the GMSB whose messengers are
singlet under the extra gauge interaction.
Then the gaugino of the strong
sector is massless from the beginning
and therefore the A-parameter is not
enhanced.
In addition to that, the soft scalar masses of the 
extra matter fields 
are as small as slepton masses.
Explicit considerations of such GMSB scenarios
will be reported elsewhere.

%%%%%%%%%%%%%%%%%%%%%%%%%%%%%
\section{Higgs coupled with a superconformal sector}
In this section we consider an explicit model
in which the Higgs field couples with a superconformal
gauge theory through a new Yukawa coupling.
First we assume the extra gauge group $G_S$ to be 
$SU(3)_{\rm SC}$,
and the color $SU(3)_C$ of the standard model
is given by
the diagonal subgroup of 
$SU(3)_{\rm SC} \times SU(3)'_C$.
We introduce vector-like matter fields charged under
$SU(3)_{\rm SC}$ so that there is an IR fixed point
with strong gauge coupling.
Then the gauge couplings of these groups are related
with $1/g^2_3 = 1/g^2_{\rm SC} + 1/g^{2'}_3$.
Therefore the gauge coupling of $SU(3)'_C$ 
is almost the same as the color gauge coupling,
$g_3 \sim g_3'$, since $g_{\rm SC}$ is as
large as $4\pi$ at the fixed point.
The gaugino mass at low energy, $M_3$, is related with
the gaugino masses $M_{\rm SC}$ and $M_3'$ of 
$SU(3)_{\rm SC}$ and $SU(3)'_C$ sectors  by 
\be
\frac{M_3}{g^2_3} = \frac{M_{\rm SC}}{g^2_{\rm SC}}
+ \frac{M_3'}{g^{2'}_3}.
\ee
It is seen that $M_3 \sim M_3'$, since $M_{\rm SC}$
is suppressed by the superconformal dynamics.

The matter contents other than the fields of the MSSM 
are as follows;
\be
\begin{array}{c|ccccc}
 & SU(3)_{\rm SC} & SU(3)'_C & SU(2)_W & U(1)_Y & R \\ \hline
\Phi & {\bf 3} & {\bf 1} &  {\bf 2} &  1/6 & -  \\
\bar{\Phi} & {\bf 3}^* & {\bf 1} &  {\bf 2} &  -1/6  & -  \\
\phi & {\bf 3} & {\bf 1} &  {\bf 1} &  2/3 & -  \\
\bar{\phi} & {\bf 3}^* & {\bf 1} &  {\bf 1} &  -2/3 & -  \\
\Omega & {\bf 3} & {\bf 3}^* &  {\bf 1} &  0 & +  \\
\bar{\Omega} & {\bf 3}^* & {\bf 3} &  {\bf 1} &  0 & + 
\end{array}
\ee
Here $R$ denotes the R-parity. 
Quarks in the MSSM belong to the fundamental
representation of $SU(3)'_C$. 
The extra matter fields are not combined into representations 
of an $SU(5)$ because of their $U(1)_Y$ charge assignment. 
We will see later that gauge coupling 
unification is slightly modified.

The matter fields charged under $SU(3)_{\rm SC}$
are vector-like and the number of the ``flavor" $N_f$ is 6.
Therefore this gauge theory belongs to
the conformal window given
by $3/2 N_c < N_f < 3 N_c$ and has an IR
attractive fixed point. 
If we neglect  the gauge couplings of 
$SU(3)'_C, SU(2)_W$ and $U(1)_Y$,
the anomalous dimensions of the vector-like matter fields are
exactly given by 
$\gamma_{\Phi}=\gamma_{\phi}
=\gamma_{\Omega}=(3N_C - N_f)/N_c = -1/2$ 
at the fixed point. 

These negative anomalous dimensions mean that
the dimensions of the charged matter fields are 
less than the canonical ones ($d = 1/2$). 
So the Yukawa term $\lambda \Phi \bar{\phi} H_u$,
which is allowed by the symmetry, 
is a dimension two operator
at the fixed point of the pure gauge theory.
Therefore this perturbation is 
relevant and the coupling $\lambda$
grows towards IR with a single power of the scale.
However the large Yukawa coupling $\lambda$ increases
the dimension of the Higgs $H_u$, and 
is expected to eventually approach a new fixed 
point $\lambda_*$.

Though existence of this new fixed point has not 
been proven yet, we may demonstrate it by applying 
one-loop anomalous dimensions in the exact 
RG equations \cite{NSVZ}.
By neglecting the MSSM couplings, the anomalous
dimensions are found to be 
\bea
\gamma_{\Phi} &=& -\frac{8}{3} \alpha_{\rm SC}
+ \alpha_{\lambda}, \nn \\
\gamma_{\bar{\phi}} &=& -\frac{8}{3} \alpha_{\rm SC}
+ 2 \alpha_{\lambda}, \nn \\
\gamma_{H_u} &=& 3 \alpha_{\lambda}, \nn \\
\gamma_{\bar{\Phi}} &=& 
\gamma_{\phi} = \gamma_{\Omega} = \gamma_{\bar{\Omega}} = 
-\frac{8}{3} \alpha_{\rm SC},
\eea
where $\alpha_{\rm SC}=g^2_{\rm SC}/8\pi^2$ and
$\alpha_{\lambda}=|\lambda|^2/8\pi^2$.
The exact beta functions are given in terms of these
anomalous dimensions as
\bea
\frac{d \alpha_{\rm SC}}{d \ln \mu} &=&
-\frac{\alpha_{\rm SC}^2}{1 - 3 \alpha_{\rm SC}}
\left(
3 + \gamma_{\Phi} + \frac{1}{2} \gamma_{\bar{\phi}}
\right. \nn \\
& & 
\left.
\gamma_{\bar{\Phi}} + \frac{1}{2} \gamma_{\phi}
+ \frac{3}{2} \gamma_{\Omega} 
+ \frac{3}{2} \gamma_{\bar{\Omega}}
\right), 
\label{betag}\\
\frac{d \alpha_{\lambda}}{d \ln \mu} &=&
\alpha_{\lambda} \left(
\gamma_{\Phi} + \gamma_{\bar{\phi}} + \gamma_{H_u}
\right).
\label{betalambda}
\eea
Immediately the fixed point couplings are found to 
be 
\bea
&{\rm (A)}:& 
(\alpha_{\rm SC}^*, \alpha_{\lambda}^*) = (3/16, 0), 
\label{fpA} \\
&{\rm (B)}:&
(\alpha_{\rm SC}^*, \alpha_{\lambda}^*) = (27/128, 3/16),
\label{fpB}
\eea
which are marked in Fig.~1.
The RG flows are also shown in Fig.~1. It is clearly seen
that the fixed point (B) is indeed IR attractive.
Now we suppose that the theory is given near the 
fixed point (A) and 
comes close to the IR attractive fixed point (B) 
at scale $\Lambda_{\rm SC}$ 
that is not much larger than the 
Electro-weak scale, say $10 \sim 100$TeV \cite{KT}.

\begin{figure}[htb]
\includegraphics[width=8cm]{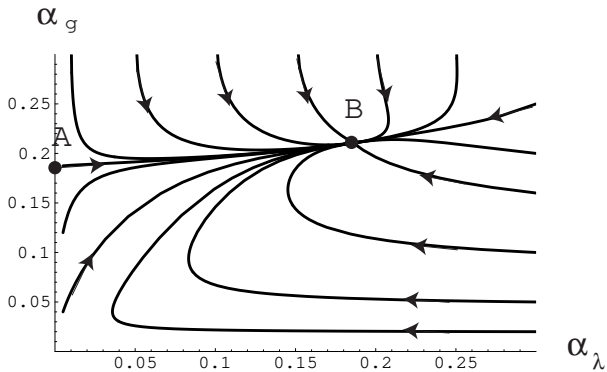}
\caption{\label{fig1} 
The RG flows obtained by solving 
Eqs.~(\ref{betag}, \ref{betalambda}) are shown. The arrows 
indicate the direction toward IR. The points A and B represent
the fixed points (\ref{fpA}) and (\ref{fpB})
respectively.}
\end{figure}

There are other relevant operators than 
$\lambda \Phi \bar{\phi} H_u$ at the fixed point (A).
We include these operators into the superpotential 
and suppose that it is given at the IR attractive
fixed point as 
\bea
W &=& Y_t Q_3 \bar{u}_3 H_u +  \lambda_* \Phi \bar{\phi} H_u 
+ \kappa_* Q_3 \Omega \bar{\Phi} 
+ \kappa'_* \phi \bar{\Omega} \bar{u}_3 \nn \\
& &
+ \mu H_u H_d + M_{\Phi} \Phi \bar{\Phi} 
+ M_{\phi} \phi \bar{\phi} + M_{\Omega} \Omega \bar{\Omega}.
\label{superpotential}
\eea
Here the supersymmetric mass terms are also 
introduced.
We do not consider the Yukawa coupling with $H_d$ for
simplicity.
When we neglect the gauge couplings $g_3'$ as well
as $g_2$ and $g_1$, the existence of the IR attractive
fixed point with large Yukawa couplings 
$\lambda_*, \kappa_*, \kappa'_*$ may be
shown by using the anomalous dimensions
approximated in the one-loop level.
Then we may find an IR attractive fixed point 
and the explicit couplings are given by
$(\alpha_{\rm SC}^*, \alpha_{\lambda}^*,
\alpha_{\kappa}^*, \alpha_{\kappa'}^*)
\simeq (0.38, 0.34, 0.25, 0.29)$,
where $\alpha_{\kappa}=|\kappa|^2/8\pi^2$.
Of course such a perturbative analysis is 
not so trustworthy for these large couplings, 
though it may offer us an indication of
the fixed point.

The anomalous dimensions for the extra matter fields
are modified in the presence of these Yukawa terms.
Then the anomalous dimension of Higgs is given by
\be
\gamma_{H_u} = - (\gamma_{\Phi} + \gamma_{\phi}),
\ee
which is expected to be about one.
The mass parameter $\mu$ is also suppressed as 
\be
\mu(\Lambda) = 
\left( \frac{\Lambda}{\Lambda_{\rm SC}} \right)^{\gamma_{H_u}}
\mu(\Lambda_{\rm SC}),
\ee
while the mass parameters 
$M_{\Phi}, M_{\phi}, M_{\Omega}$ are enhanced.
Therefore the decoupling scale of the superconformal sector
is much larger than $\mu$.
In this model, not only Higgs but also quarks acquire
positive anomalous dimensions, therefore the Yukawa couplings
$Y_t$ is suppressed rather strongly.

The spontaneous breakdown of the gauge symmetry 
$SU(3)_{\rm SC} \times SU(3)'_C$ to the diagonal subgroup 
$SU(3)_C$ occurs, 
if the bi-fundamental matter fields acquire VEVs as
\be
\langle \Omega^A_a \rangle =  \omega \delta^A_a,~~~~
\langle \bar{\Omega}_A^a \rangle =  \bar{\omega}  \delta_A^a,
\ee
where $A, a = 1, 2, 3$ are indices of the 
fundamental representations.
These VEVs also bring about mixing between the 
quark superfields and
the superconformal matter fields simultaneously.
By substituting these VEVs into the superpotential, 
the mass terms are modified effectively as
\be
W \sim 
M_{\Phi} \left(
\Phi + \frac{\kappa_* \omega}{M_{\Phi}} Q_3
\right)\bar{\Phi}
+ M_{\phi}\phi 
\left(
\bar{\phi} + \frac{\kappa'_* \bar{\omega}}{M_{\phi}} \bar{u}_3
\right).
\ee
It is noted that the decoupling modes are 
given by a linear combination
with the original quark fields. Therefore, 
a new Yukawa term is induced in the effective 
superpotential after decoupling.
Thus the effective top Yukawa coupling is found to be
\be
Y_t^{\rm eff} \sim \lambda_* 
\left(\frac{\kappa_* \omega}{M_{\Phi}} \right) 
\left(\frac{\kappa'_* \bar{\omega}}{M_{\phi}} \right).
\ee
The top Yukawa coupling of 
$Y_t^{\rm eff} \sim 1$ may be realized if 
$\omega/M_{\Phi} \sim \bar{\omega}/M_{\phi} 
\sim 1/\kappa_* \sqrt{\lambda_*}$.
Here the fixed point couplings $\lambda_*, \kappa_*$
are expected to be $2\pi/3$ as a rough estimation.
Thus the large top Yukawa coupling may be reproduced
by assuming $\omega$ to be given in mass scale of
the extra matter fields.

Next we consider the soft scalar mass of Higgs
obtained at the decoupling scale from the 
superconformal sector.
The exact beta functions \cite{NSVZ} and 
the spurion method \cite{yamada,HS,JJ,kazakov,KKZ,AGLR}
enable us to show that the gaugino mass and 
the A-parameters in a superconformal gauge theory
decrease in proportion to certain
powers of the scale towards IR \cite{Karch,sumrule}.
Therefore these supersymmetry breaking parameters 
are almost vanishing at the scale of $\Lambda_{\rm SC}$,
where the Yukawa coupling $\lambda$ reaches near the IR
fixed point.
Since the coupling $Y_t$ also becomes
negligible below the scale of $\Lambda_{\rm SC}$, 
the RG equations for 
$m^2_{\Phi}, m^2_{\bar{\phi}}, m^2_{H_u}$ 
are just identical to Eqs.~(\ref{topRG}) at one-loop
level.
It is sufficient to replace  
$m^2_{Q_3}, m^2_{\bar{u}_3}$ to
$m^2_{\Phi}, m^2_{\bar{\phi}}$ and
$Y_t, M_3, A_t$ to 
$\lambda, M_{\rm SC}, A_{\lambda}$ respectively
in Eqs.~(\ref{topRG}).
Here, however, the soft breaking parameters 
$M_{\rm SC}, A_{\lambda}$ have been sufficiently suppressed 
already at $\Lambda_{\rm SC}$.
Therefore these soft masses, especially 
$m^2_{H_u}$, are not enhanced below $\Lambda_{\rm SC}$,
as is discussed in the previous section.
Consequently if the scalar masses 
$m^2_{\Phi}, m^2_{\bar{\phi}}, m^2_{H_u}$ 
are all small at scale of $\Lambda_{\rm SC}$,
then $m^2_{H_u}$ becomes also small at
low energy.

Now we suppose that the theory stays near the fixed 
point (A) given in Eq.(\ref{fpA}) above a certain
scale of $\Lambda'_{\rm SC} > \Lambda_{\rm SC}$, 
and the coupling $\lambda$ is tiny.
There the Higgs field is separated from the strongly
interacting sector completely.
In the absence of the large top Yukawa coupling $Y_t$,
the dominant radiative corrections to the Higgs
mass parameters are given by the $SU(2)_W$ gauge
interaction, which are irrelevant to fine-tuning.
On the other hand only the gauge interaction
is strong in the superconformal sector
and the extra vector-like
matter fields are subject to the same corrections.
Then, if the scalar masses of
these fields are given to be universal at the
fundamental scale, 
the scalar masses are reduced through the superconformal dynamics
\cite{Karch,sumrule,KT}.
Thus the scalar masses $m^2_{\Phi}$ and 
$m^2_{\bar{\phi}}$ as well as the gaugino mass
$M_{\rm SC}$ have been suppressed at the scale 
$\Lambda'_{\rm SC}$.

Finally let us consider the RG behavior of the 
soft scalar masses at the transition region 
from fixed point (A) to (B) in more details.
It is noted that the A-parameter $A_{\lambda}$ 
is not suppressed before approaching the
fixed point (B).
Meanwhile the Yukawa coupling $\lambda$ grows rapidly
and exceed O(1) quickly.
Therefore we may wonder that the A-parameter
affects RG behavior of the soft scalar masses
significantly.

In order to see this, we solve the coupled
RG equations for 
$m^2_{\Phi}, m^2_{\bar{\phi}}, m^2_{H_u}$ 
and $A_{\lambda}$.
The explicit equations may be written down 
immediately by applying 1-loop anomalous dimensions
to the exact formulae
\cite{yamada,HS,JJ,kazakov,KKZ,AGLR}.
{}First the RG equation for 
$A_{\lambda}$ is found to be
\be
\frac{d A_{\lambda}}{d \ln \mu} =
\frac{16}{3}\alpha_{\rm SC} M_{\rm SC}
+ 6 \alpha_{\lambda} A_{\lambda},
\label{rg-Al}
\ee
where $\alpha_{\rm SC} = g^2_{\rm SC}/8\pi^2$
and $\alpha_{\lambda} = |\lambda|^2/8\pi^2$
again.
It is seen that 
the A-parameter $A_{\lambda}$ is suppressed 
rapidly as $\lambda$ grows
in the absence of the gaugino mass $M_{\rm SC}$.
The RG equations for the scalar masses may be
obtained similarly and are found to be
\bea
\frac{d m^2_{\Phi}}{d \ln \mu} &=&
-\frac{8}{3} \alpha_{\rm SC}
\left( 2 |M_{\rm SC}|^2 + \Delta_g \right) \nn \\
& & + \alpha_{\lambda}
\left( |A_{\lambda}|^2 + \Sigma \right), \\
\frac{d m^2_{\bar{\phi}}}{d \ln \mu} &=&
-\frac{8}{3} \alpha_{\rm SC}
\left( 2 |M_{\rm SC}|^2 + \Delta_g \right) \nn \\
& & + 2 \alpha_{\lambda}
\left( |A_{\lambda}|^2 + \Sigma \right), \\
\frac{d m^2_{H_u}}{d \ln \mu} &=&
3 \alpha_{\lambda}
\left( |A_{\lambda}|^2 + \Sigma \right),
\label{rg-mass}
\eea
where $\Delta_g$ and $\Sigma$ are given in terms
of the scalar masses as 
\bea
\Delta_g &=& 
\frac{\alpha_{\rm SC}}{1 - 3 \alpha_{\rm SC}}
\left(
3|M_{\rm SC}|^2 - m^2_{\Phi} 
- \frac{1}{2} m^2_{\bar{\phi}}
\right), \\
\Sigma &=&
m^2_{\Phi} + m^2_{\bar{\phi}} + m^2_{H_u}.
\label{factors}
\eea
Here we neglected dependence on other 
scalar masses in $\Delta_g$ since these are
suppressed. 

In Fig.~2 and Fig.~3 an example of the solutions
to Eqs.~(\ref{rg-Al}$-$\ref{rg-mass}) coupled 
with Eqs.(\ref{betag}) and (\ref{betalambda})
is shown.
In these figures the parameter $t$ is related with
the renormalization scale $\mu$ by 
$t=\log_{10}(\mu/\Lambda'_{\rm SC})$.
At $t=0$,
the scalar masses $m^2_{\Phi}$ and $m^2_{\bar{\phi}}$
as well as the gaugino mass $M_{\rm SC}$
are supposed to be suppressed.
On the other hand, the scalar mass of Higgs $|m_{H_u}|$ is 
not suppressed by superconformal dynamics,
though it may be smallish.
Therefore we examined the case that $A_{\lambda}$ is
given somewhat larger than $|m_{H_u}|$ 
(to be explicit, $m^2_{H_u}/A^2_{\lambda} = 0.1$)
at $t=0$.
We also set $m^2_{\Phi} = m^2_{\bar{\phi}} =0$
at $t=0$ for simplicity.
In Fig.~2 running aspect of $A_{\lambda}$ is shown 
in the transition between the fixed points.
Indeed the A-parameter is suppressed rapidly
as the Yukawa coupling $\alpha_{\lambda}$
approaches  the IR fixed point (B).
The solutions for scalar masses
$m^2_{\Phi}, m^2_{\bar{\phi}}, m^2_{H_u}$ 
are shown in Fig.~3.
We find that $m^2_{H_u}$ is 
reduced somewhat by the A-parameter, but the
correction is not considerable,
even though $A_{\lambda}$ is relatively 
large at $\Lambda'_{\rm SC}$. 

\begin{figure}[htb]
\includegraphics[width=0.45\textwidth]{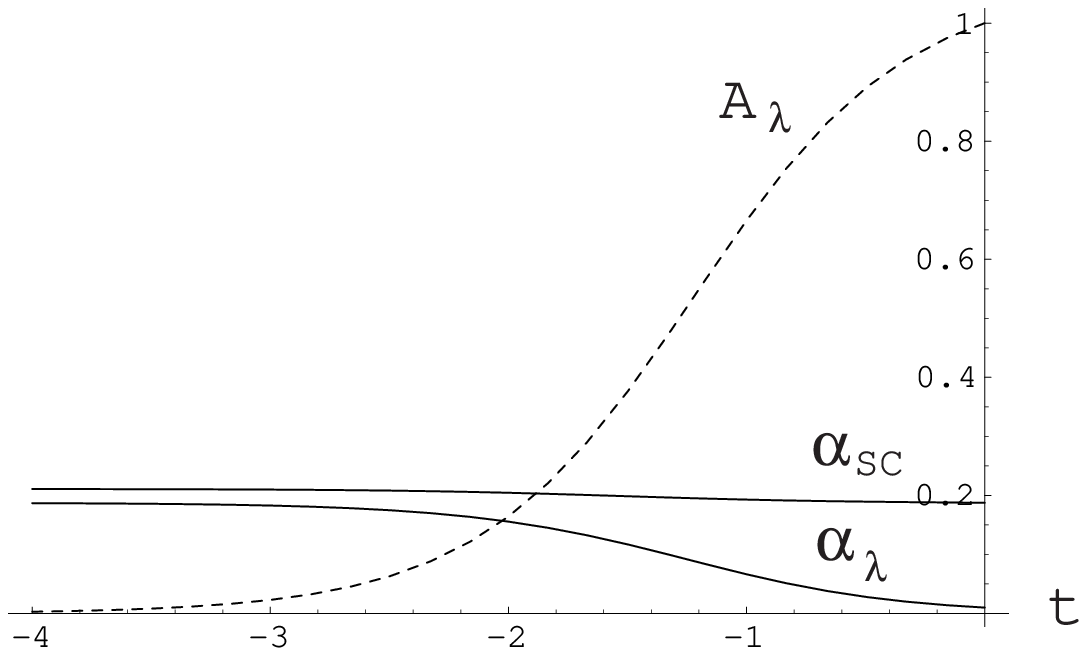}
\caption{\label{fig2} Running behavior of
$A_{\lambda}$ as well as the couplings,
$\alpha_{\rm SC} = g^2_{\rm SC}/8\pi^2$ and
$\alpha_{\lambda} = |\lambda|^2/8\pi^2$,
in the transition region between the fixed points
(A) and (B) is shown. 
The renormalization scale $\mu$ is represented 
by $t= \log_{10}(\mu/\Lambda'_{\rm SC})$.
These lines are obatined by solving 
Eqs.~(\ref{betag}), (\ref{betalambda}) and 
(\ref{rg-Al}).
}
\includegraphics[width=0.45\textwidth]{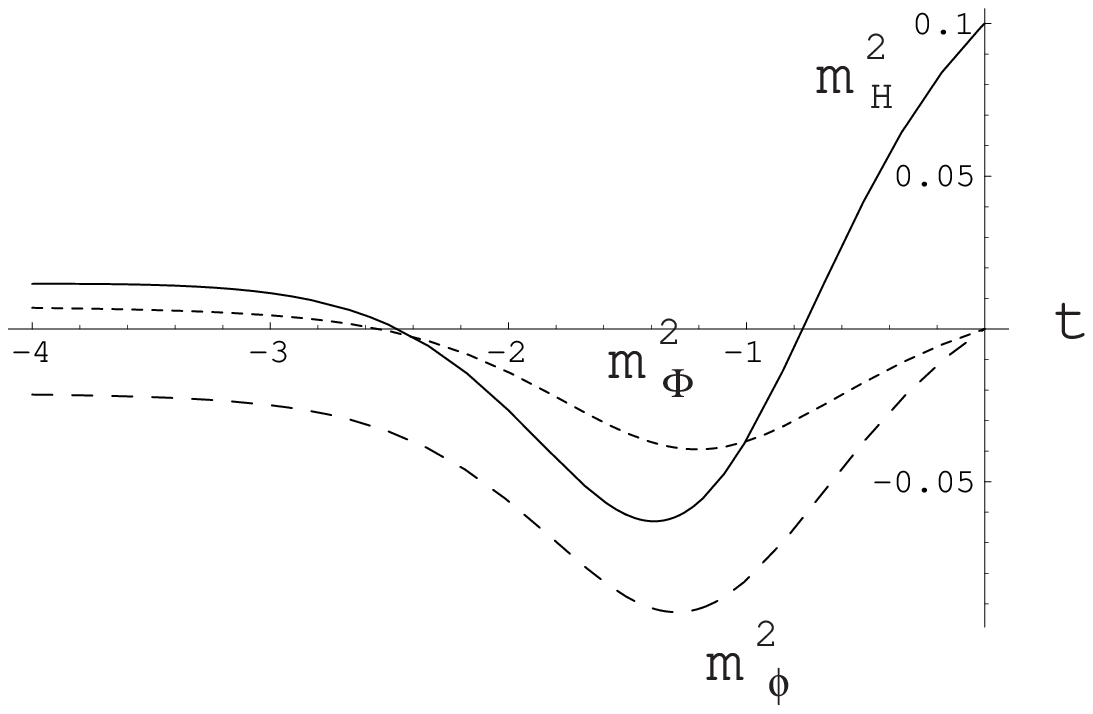}
\caption{\label{fig3} Running behavior of
soft scalar masses $m^2_H$, $m^2_{\Phi}$ and
$m^2_{\bar{\phi}}$ is shown. 
The renormalization scale $\mu$ is represented 
by $t= \log_{10}(\mu/\Lambda'_{\rm SC})$.
We set $m^2_{H_u}/A^2_{\lambda} = 0.1$
at $t=0$ as an example.
These lines are obatined by solving 
Eqs.~(\ref{rg-Al}$-$\ref{rg-mass}) coupled 
with Eqs.(\ref{betag}) and (\ref{betalambda}).
}
\end{figure}

So far we have not taken into account the $SU(3)'_C$ effects 
to the soft scalar masses and have assumed 
that all scalar masses of the matter fields charged under
$SU(3)_{\rm SC}$  are suppressed by superconformal 
dynamics \cite{Karch}.
However the extra matter fields $\Omega$ and $\bar{\Omega}$,
whose VEVs bring about the symmetry breaking and 
the mixing between the original top quark and superconformal 
matter fields,
are charged under $SU(3)'_C$.
So one may wonder if $m^2_{\Omega}$ and 
$m^2_{\bar{\Omega}}$ are enhanced by the $SU(3)'_C$
correction just like squark masses, since 
mass of the $SU(3)'_C$ gaugino is supposed to become
large at low energy.
In addition, $m^2_{\Phi}$ and $m^2_{\bar{\phi}}$ are not just
suppressed but are also enhanced as much as 
$m^2_{\Omega}$, because these scalar masses should
satisfy the IR sum rule \cite{sumrule}.

This may be avoided as follows.
It is noted that the symmetry allows interaction 
among $\Omega$ given by
\be
\Delta W = \eta_*~\epsilon_{ABC}~\epsilon^{abc}
\Omega^A_a~\Omega^B_b~\Omega^C_c,
\ee
in the superpotential. 
Then the IR sum rule tells us $m^2_{\Omega} \rightarrow 0$
at low energy.
In practice, $m^2_{\Omega}$ does not vanish because of the
$SU(3)'_C$ corrections.
However we note that the logarithmic correction does 
disappear \cite{KT}.
This may be demonstrated by using the one-loop RG equation for 
$m^2_{\Omega}$, which is given by
\be
16 \pi^2 \frac{d m^2_{\Omega}}{d \ln \mu}
= 4|\eta_*|^2 m^2_{\Omega} - \frac{32}{3} g^2_3 |M_3|^2.
\ee
If we use $|\eta_*| \sim 4\pi$ by naive 
dimensional analysis, then the soft mass converges as
\be
m^2_{\Omega} \rightarrow \frac{1}{6 \pi^2} g^2_3 |M_3|^2.
\ee
Thus $m^2_{\Omega}$ becomes small enough at the scale of
$\Lambda'_{\rm SC}$.
Then the IR sum rule
guarantees that the soft scalar masses 
$|m^2_{\Phi}|$ and  $|m^2_{\bar{\phi}}|$ 
are as small as $m^2_{\Omega}$.
Thus although the mass of $SU(3)'_C$ gaugino may be large,
the soft scalar masses in the 
superconformal sector can be suppressed.

%%%%%%%%%%%%%%%%%%%%%%%%%%
\section{Some phenomenological aspects}
First we consider the gauge coupling unification.
Introduction of the extra matter fields charged under
the SM gauge group alters the gauge beta functions.
As was mentioned before, the extra matter fields are not
combined into representations of $SU(5)$.
Therefore the Weinberg angle is shifted from the
value obtained in the MSSM, which fits in the
experimental data.

The running gauge couplings 
$\alpha_a=g^2_a/4\pi (a=1,2,3)$ for 
the SM gauge groups are given explicitly at one-loop 
level;
\begin{eqnarray}
\alpha^{-1}_a(\mu) &=& \alpha^{-1}_a(M_Z)  
+ \frac{b^{\rm (low)}_a}{2 \pi} \ln \frac{M_{\Phi}}{M_Z} 
\nn \\
& &
+ \frac{b^{\rm (SC)}_a}{2 \pi} \ln \frac{\Lambda_{\rm SC}}{M_{\Phi}} 
+ \frac{b^{\rm (high)}_a}{2 \pi} \ln \frac{\mu}{\Lambda_{\rm SC}},
\label{MSSMgauge}
\end{eqnarray}
where $M_{\Phi}$ stands for the decoupling scale of the
extra matter fields. We also suppose the scale $\Lambda'_{\rm SC}$
to be rather close to $\Lambda_{\rm SC}$ and do not consider
running between these scales.
To be explicit we use $M_{\Phi} = 1$TeV and 
$\Lambda_{\rm SC} = 10$TeV in the calculations below.

Now we calculate the beta function 
coefficients $b_a$ in each energy region.
In the present model only the Higgsino and probably 
the right-handed sleptons are supposed to be light
among superpartners of the SM fields.
Then $b^{\rm (low)}_a$ may be evaluated as
\be
(b^{\rm (low)}_3, b^{\rm (low)}_2, b^{\rm (low)}_1) 
= (-7, -15/6, 51/10).
\ee
Though the running below $M_{\Phi}$ is rather model
dependent and uncertain, the contribution is not 
important anyway.

Next we consider the region where the MSSM coupled with
the superconformal sector.
We suppose that not only the coupling $\lambda$
but also $\kappa$ and $\kappa'$ are reaching their
IR fixed points below $\Lambda_{\rm SC}$ in the
superpotential given by (\ref{superpotential}).
There the anomalous dimensions of the fields 
charged under $SU(3)_{\rm SC}$,
$\gamma_{\Phi}, \gamma_{\bar{\Phi}}$ and so on,
are $-1/2$ approximately.
Meanwhile $\gamma_{H_u}, \gamma_{Q_3}$ and $\gamma_{\bar{u}_3}$
may be evaluated to be about one.
The gauge beta function coefficients are given in general
by
$
b_a = - b^{\rm (MSSM)}_a  + \sum_i (1/2)
(1 - \gamma_{\phi_i}),
$
where the chiral fields $\phi_i$ belong to the
fundamental representation.
Therefore we may evaluate $b^{\rm (SC)}_a$ as
\be
(b^{\rm (SC)}_3, b^{\rm (SC)}_2, b^{\rm (SC)}_1) 
\sim (0, 7/2, 81/10).
\ee
Above the scale $\Lambda_{\rm SC}$ 
(or $\Lambda'_{\rm SC}$), the couplings
of the superconformal sector are also supposed to be
small.
Therefore the coefficients may be evaluated by one-loop
approximation and are found to be
\be
(b^{\rm (high)}_3,b^{\rm (high)}_2,b^{\rm (high)}_1) 
= (0, 4, 42/5).
\ee

By using these coefficients in the gauge couplings
Eqs.~(\ref{MSSMgauge}), we may examine
the aspect of their unification.
{}For example it is seen that 
$\alpha_3$ and 
$\alpha_1$ cross each other at the scale about $10^{18}$GeV.
It may be interesting that this is so 
high as the Planck scale or the string scale.
If we assume that $\alpha_2$ coincides with other gauge
couplings at this scale, then the Weinberg angle is
found to be
$\sin^2 \theta_W \approx 0.24$, which is still very close
to the realistic value.

Next we also examine low energy spectrum of the
superparticles in this model.
Specifically we consider the mSUGRA scenario with the
universal gaugino mass $M_{1/2}$ and the universal
scalar mass $m_0$, which is now assumed to be 
very small at the  Planck scale.
The running behavior of the gauge couplings are
rather different from that in the MSSM. 
Therefore the soft parameters obtained at low energy
are distinct from those in the ordinary mSUGRA scenario.

It will be enough to take account of the corrections above
the scale of $\Lambda_{\rm SC}$ at the one-loop level.
Since the gaugino masses satisfy the relation,
$
M_3(\mu) : M_2 (\mu) : M_1 (\mu) = \alpha_3(\mu) : 
\alpha_2 (\mu) : \alpha_1 (\mu),
$
their ratios are the same as in the mSUGRA
case;
\be
M_2 \approx 0.29 M_3, \qquad 
M_1 \approx 0.14 M_3.
\ee
Contrary to this, the squark masses are considerably
enhanced by the $SU(3)_C$ corrections, since
$\alpha_3$ is not reduced at higher energy scale.
Explicitly the correction may be given as
\be
\Delta m^2_{\tilde q} = \frac{32}{12 \pi}
\alpha_3 |M_3|^2 \ln 
\frac{M_{\rm Pl}}{\Lambda_{\rm SC}},
\ee
from which the squark masses at low energy are 
expected to be
\be
m_{\tilde Q} \sim m_{\tilde q} \sim 1.6 M_3.
\ee
Similarly we may obtain other soft mass parameters
as
\be
m_{\tilde L} \sim 0.5 M_3, \quad
m_{H_d} \sim 0.5 M_3, \quad
m_{\tilde e} \sim 0.28 M_3.
\ee
We note that the soft masses in this model 
are relatively heavier than those 
in the conventional scenarios.
In the presence of $m_0$ at the Planck scale, these are
raised up more.

On top of that the most characteristic feature of 
our model is that Higgs and Higgsino fields are 
much lighter than squark and sleptons.
The soft scalar mass $m^2_{H_u}$ receives radiative corrections
only below the decoupling scale $M_{\Phi}$ and is expected to be
one order smaller than $m^2_{\tilde{t}}$.
The $\mu$-parameter is also suppressed. 
This spectrum is favorable in the following respects.
One is the neutralino relic abundance \cite{darkmatter},
which has been constrained precisely by WMAP.
In the most parameter region of mSUGRA  the lightest 
neutralino is Bino-like, and the relic abundance
constraint leads to stringent upper bound on scalar
and/or gaugino masses.
In our model the lightest supersymmetric particle (LSP)
is given as a gaugino-Higgsino mixture or even as a
Higgsino dominant component.
It has been known that the relic abundance is well 
explained in the case of such spectra, which is similar to
the focus point region \cite{focus,darkmatter}.

Another advantageous feature is on the stability of the
MSSM scalar potential. The mass parameters in the MSSM 
must satisfy constraints against charge and/or color breaking 
and unbounded-from-below.
The most serious unbounded-from-below direction 
of the MSSM scalar potential involves the Higgs and 
slepton fields, i.e. the so-called UFB-3 
direction \cite{UFB}.
If the quantity 
$m^2_{H_u} + m_{\tilde L}^2$ turns out to be negative,
then the potential becomes unstable along the UFB-3 
direction.
In our scenario, the Higgs soft mass is suppressed, and
therefore the UFB-3 bound can be relaxed.

Lastly we consider also the constraints by precision 
measurement of the EW theory.
In our scenario it is essential to incorporate the mixing
of the top quark with a heavy extra field in order to generate
heavy top mass.
We come across a similar situation in the so-called
top quark see-saw models \cite{topseesaw}.
In these models, a vector-like pair of weak singlet 
fermions $(\chi_L, \chi_R)$ are introduced and 
mixed mass terms with top quarks as well as mass
of themselves are generated dynamically.
Then the stringent constraints to the isospin breaking 
$\delta \rho$ and the Z-boson decay width 
$R_b=\Gamma[Z \rightarrow b \bar{b}]
/\Gamma[Z \rightarrow \mbox{\rm hadrons}]$
are found to give lower bounds for the dynamical masses
\cite{topseesaw}.
After EWSB, similar mixed mass terms 
with extra matter fields appear also in our model.
However masses of the extra matters are supposed to be 
more than a few TeV, while mixed masses are as 
small as the VEV of EWSB. 
Therefore the mixing effects are small enough to
satisfy the constraints for  $\delta \rho$ and $R_b$.

\section{Conclusion and discussion}
In this article we considered a scenario ameliorating 
the supersymmetric little hierarchy problem in the
MSSM. 
We concentrated particularly on models with heavy gluinos and
squarks as given in mSUGRA and GMSB scenarios.
In order to suppress radiative corrections to
Higgs mass parameters, we assumed that top Yukawa 
coupling is small above TeV.
The large top quark mass is effectively generated
through top quark mixing with strongly coupled sector 
around TeV scale.
Accordingly the Higgs field has a large Yukawa coupling with the
extra matter fields instead of the top quark. 
Nevertheless  corrections to the soft scalar mass of Higgs 
can be made suppressed.

To be explicit, we presented a model in which the Higgs field
couples with a superconformal gauge theory.
Then it is shown that Higgs is sequestered from large
supersymmetry breaking effects.
Moreover the large anomalous dimension of the Higgs field
suppresses the $\mu$-term.
Thus Higgs and Higgsino are light compared with
squarks and sleptons.
The GMSB scenarios with messengers that are
neutral under the strong gauge group will also be considerable
and supposed to present a similar sparticle spectrum.

In the model building we just assumed the symmetry
breaking of $SU(3)_{\rm SC} \times SU(3)'_C$
to occur also around TeV scale. 
Though it is interesting to construct 
an explicit model that implements this symmetry breaking, 
we leave it for future study.
We also assumed that only top quarks acquire mixing terms
with the extra matter fields. 
However this may be related with the question why only top
quark mass is prominently large.
In this respect it may be interesting to extend the model
to incorporate three generations of quarks,
and to consider the quark mass matrix.

Lastly some comments on the other corrections induced
by the extra matter fields are in order.
Recently Babu {\it et. al.} \cite{Babu} considered 
similar models to ours, though they were not concerned
with soft scalar masses and their models are 
rather weakly coupled.
They examined the cases that additional contributions
to the quartic coupling of Higgs bosons in the MSSM
are induced by the loop effect of extra matter fields
coupled to Higgs fields.
Then the mass of the lightest neutral
Higgs boson may be  raised up considerably.
In our model the same kind of correction also exists
and is given roughly by
\be
\Delta m^2_{h^0} \sim
\frac{3}{8\pi^2} \left(
-m_Z^2 \cos^2 2\beta \lambda_*^2
+ 2 v^2 \sin^4 \beta \lambda_*^4
\right) t_1,
\ee
where $t_1 = \ln (1 + m^2_{\Phi}/M^2_{\Phi})$
and we used $M_{\bar{\phi}} = M_{\Phi}$ and
$m^2_{\bar{\phi}} = m^2_{\Phi}$ for
simplicity.
However the supersymmetric mass $M_{\Phi}$ is
now taken to be fairly large, while the soft mass
$m_{\Phi}$ is suppressed. 
Therefore enhancement of the lightest Higgs boson
mass is found to be very small in the present model.

We add remarks on the
decoupling effects to the soft supersymmetry
breaking parameters.
We note that B-parameters accompanied with the
mass terms of the extra matter fields exist and are not
suppressed by the superconformal dynamics.
Even the B-parameters are induced through corrections
with the gaugino for $SU(3)_{\rm SC}$, 
since mass of the gaugino is not small at very high 
energy scale.
Then soft supersymmetry breaking parameters 
in the MSSM receive threshold
corrections due to  the B-parameters when the
extra matter fields decouple at the mass scale 
of $M_{\Phi}$.
{}For gaugino masses, the threshold corrections
are similar to the gauge mediation effect.
What we especially concern ourselves about is 
the effect to the soft scalar mass of Higgs, 
$m^2_{H_u}$.
We can evaluate such threshold corrections
in a mannar outlined in Ref.~\cite{Nakano}.
Then we find that  the threshold 
correction to $m^2_{H_u}$ may be evaluated as 
\be
\Delta m^2_{H_u}
\sim - |B|^2 \Delta \left(
\frac{d \gamma_{H_u}}{d \ln \mu}
\right),
\ee
where $\Delta(*)$ represents the difference generated 
through the decoupling.
Here the scale dependence of the anomalous 
dimension of the Higgs field is brought about
by the SM gauge interactions in the leading
order. Therefore, the threshold corrections 
to $m^2_{H_u}$ are found to be
insignificant for the present problem.

\section*{Acknowledgements}
T.~K.\ is supported in part by the Grants-in-Aid for 
Scientific Research
%(No.~14540256)
(No.~16028211) 
and the Grant-in-Aid for the 21st Century COE
``The Center for Diversity and Universality in Physics"
from the Ministry of Education, Science, Sports and 
Culture, Japan.
H.~N.\ and H.~T.\ 
are supported in part by the Grants-in-Aid for 
Scientific Research
%No.~14540256 and 
(No.~16540238 and No.~13135210, respectively)
from the Ministry of Education, Science, Sports and 
Culture, Japan.

\end{document}